\documentstyle[aps,epsfig,floats]{revtex}

\begin{document}
\parskip 1mm
\draft
\twocolumn[\hsize\textwidth\columnwidth\hsize\csname@twocolumnfalse%
\endcsname
\title{Unusual scaling for pulsed laser deposition}
\author{Berit Hinnemann$^1$, Haye Hinrichsen$^2$, 
      and Dietrich E. Wolf$^1$\\[2mm]}

\address{$^1$ Theoretische Physik, Fachbereich 10,
     Gerhard-Mercator-Universit{\"a}t Duisburg,
     47048 Duisburg, Germany}
\address{$^2$ Theoretische Physik, Fachbereich 8,
     Universit{\"a}t GH Wuppertal,
     42097 Wuppertal, Germany}

\date{January 22, 2001}
\maketitle

\begin{abstract}
We demonstrate that a simple model for pulsed laser deposition
exhibits an unusual type of scaling behavior for the
island density in the submonolayer regime. This quantity is 
studied as function of pulse intensity and deposition time.
We find a data collapse for the {\em ratios of the logarithms}
of these quantities, whereas conventional scaling as observed
in molecular beam epitaxy involves ratios of powers.
\end{abstract}

]
%


Many systems in equilibrium and nonequilibrium statistical physics
exhibit power-law scaling. This means that a system with an 
observable, $M$, depending for example on two parameters, 
$z_1$ and $z_2$, looks the same, if the units of $M$, $z_1$ 
and $z_2$ are rescaled by certain factors which are related 
to each other by power-laws. Such a scaling transformation
can be written as
\begin{equation}
\label{UsualScaling}
z_1 \rightarrow \Lambda z_1, \quad z_2 \rightarrow \Lambda^{\beta}
z_2, \quad M \rightarrow \Lambda^{\alpha} M\,,
\end{equation}
where $\Lambda$ is a scaling parameter and
$\alpha,\beta$ are certain exponents. Eq.~(\ref{UsualScaling}) implies
the scaling form
\begin{equation}
\label{UsualScalingForm}
M(z_1,z_2) = z_1^{\alpha} \, f(z_2/z_1^{\beta})\,,
\end{equation}
where $f$ is a scaling function depending on
a scale-invariant argument.
This type of scaling can be observed in a 
vast variety of applications, including equilibrium 
critical phenomena \cite{Amit,Liggett}, growth processes 
\cite{EdwardsWilkinson,FamilyVicsek,KPZ,KrugSpohn,BarabasiStanley,Krug}, 
driven diffusive systems \cite{SchmittmannZia}, as well as phase
transitions far from equilibrium \cite{MarroDickman,Hinrichsen}. 

The main theoretical interest in power-law scaling stems from the
fact that the long-range properties of such systems are universal,
i.e., they are determined by certain symmetry properties and do 
not depend on microscopic details of the dynamics. This allows one
to categorize phase transitions and growth phenomena into universality
classes. Typically each universality class is associated with a
certain set of values of the critical exponents. Moreover, the
functional form of the scaling function $f$ turns out to be universal
as well.

As an example for power-law scaling, which we are going to
contrast with a different type of scaling in this paper, 
let us consider the following well-known simple model of 
molecular beam epitaxy (MBE): A particle beam deposits atoms onto a flat
substrate at a flux $F$ (atoms per unit area per unit time). 
The atoms diffuse on the substrate 
with a surface diffusion constant $D$ until they meet
another adatom, in which case they form a stable and immobile
nucleus of a two-dimensional island on the surface, or until they
attach irreversibly to the edge of an already existing island. If they
reach the edge diffusing on top of the island, they go down and attach
to the edge with the same probability as if they arrived there from
the lower terrace, i.e., Ehrlich-Schwoebel barriers are not taken into
account. 

The observable examined in this paper is the time-dependent nucleation
density, $n$, i.e., the number of nucleation events per unit area in 
the first layer integrated over time. 
Obviously, the nucleation density is a fundamental  quantity  
characterizing the island morphology as it indicates how 
many islands are formed. By definition, the nucleation density 
increases monotonically with time and saturates at a constant value
when the first monolayer is completed.

In MBE the two parameters $D$ and $F$ can be used to construct a 
characteristic length 
\begin{equation}
\label{MBE_ScalingLength}
\ell_0 = (D/F)^{1/4}.
\end{equation}
When the nucleation density reaches the value $1/\ell_0^2$, the rate
of nucleation events decreases drastically since it becomes
more likely that an adatom attaches to an already existing 
island instead of meeting another adatom and forming a nucleus.
In terms of the coverage $\Theta = F t$, i.e., the total number of
deposited atoms per unit area, the time dependence of $n$ 
is known to obey the scaling form
\begin{equation}
\label{MBE_ScalingForm}
n(\ell_0, \Theta) = \ell_0^{-2} \, f(\Theta \ell_0^2) \,.
\end{equation}
As shown in Ref.~\cite{Tang93}, 
the scaling function $f$ behaves as
\begin{equation}
\label{MBE_AsymptoticScaling}
f_1(z) \propto \left\{
\begin{array}{ll}
z^3 & \text{ for } 0 \leq z \ll 1 \\
z^{1/3} & \text{ for } 1 \ll z \ll \ell_0^2/a^2
\end{array}
\right. \,,
\end{equation}
where $a$ is the lattice constant, which will be set to unity.
Thus, MBE exhibits standard power-law scaling as described
by Eqs.~(\ref{UsualScaling}) and (\ref{UsualScalingForm}). 
However, in pulsed laser deposition (PLD),
an alternative method of growing thin epitaxial films, we find
that the nucleation density shows a fundamentally different type
of scaling, which in MBE is realized only approximately.

In order to work out the difference most clearly, let us assume
that the amplitudes of the asymptotic power laws in
(\ref{MBE_AsymptoticScaling}) are equal (which is approximately
the case in MBE), so that one can get rid of them by dividing 
$n(\ell_0,\Theta)$ by the nucleation density at a 
particular coverage $\Theta_0$. In the
following we choose $\Theta_0 = 1$ and define
\begin{equation}
M(\ell_0, \Theta)  = n(\ell_0, \Theta)/n(\ell_0, 1) = \ell_0^{-2/3} \,
f_2(\Theta \ell_0^2).
\end{equation}
\label{MDefinition}
The scaling function $f_2$ is obtained from $f_1$ by replacing the
proportionality in (\ref{MBE_AsymptoticScaling}) by an equal-sign.
If we furthermore suppose that the asymptotic power laws 
in~(\ref{MBE_AsymptoticScaling}) would remain valid right to the
crossover point at $\Theta \ell_0^2=1$ (which is
certainly not the case), then the scaling function would have the
additional symmetry
\begin{equation}
f_2(z^{\lambda}) = f_2^{\lambda}(z).
\label{AdditionalSymmetry}
\end{equation}
As a consequence the system would not only be invariant under the scale
transformation (\ref{UsualScaling}),
\begin{equation}
\ell_0 \rightarrow \Lambda \ell_0, \quad \Theta \rightarrow \Lambda^{-2}
\Theta, \quad M \rightarrow \Lambda^{-2/3} M,
\label{eq:8}
\end{equation}
but also under 
\begin{equation}
\ell_0 \rightarrow \ell_0^{\lambda}, \quad \Theta \rightarrow 
\Theta^{\lambda}, \quad M \rightarrow M^{\lambda}.
\label{eq:9}
\end{equation}
This is a scaling transformation for the {\em logarithms}
with all critical exponents equal to 1. According to
(\ref{UsualScalingForm}) this would imply
\begin{equation}
\ln M = (\ln \ell_0) \,\, g\bigl(\ln \Theta/\ln \ell_0\bigr).
\label{ApproximateScalingMBE}
\end{equation}
with a piecewise linear scaling function $g(z)$ whose slopes
are determined by the exponents in (\ref{MBE_AsymptoticScaling}). 

While in MBE this type of scaling
holds only in an approximate sense, we are going to show that 
PLD indeed obeys this scaling form
with a {\em quadratic} scaling function $g$. 
PLD is a growth technique in which the 
target material is ablated by a pulsed laser 
and then deposited in pulses on a substrate surface, 
i.e., many particles arrive simultaneously at the surface~\cite{ChriseyHubler}. 
Experimentally each pulse has a length of about a few nanoseconds and the 
time between two pulses is of the order of seconds. 
In order to investigate the scaling behavior of PLD, we consider 
a simple model which generalizes the standard 
model for MBE~\cite{ConferencePaper,FuturePaper}. 
In this model the duration of a pulse is assumed to be zero and the transient 
enhancement of the mobility of freshly deposited atoms is neglected. 
It is controlled by three parameters, namely,
the intensity, $I$, which is the density of particles deposited 
per pulse, the rate for diffusion of adatoms on the surface, $D$, and
the average flux of incident particles per site, $F$. 
Since there is no edge diffusion, the islands grow in a 
fractal manner before they coalesce.
A similar model with a finite pulse length was studied previously by
Jensen, Niemeyer, and Combe~\cite{JensenNiemeyer,CombeJensen}.

\begin{figure}
\begin{center}
\vspace{-1mm}
\epsfig{file=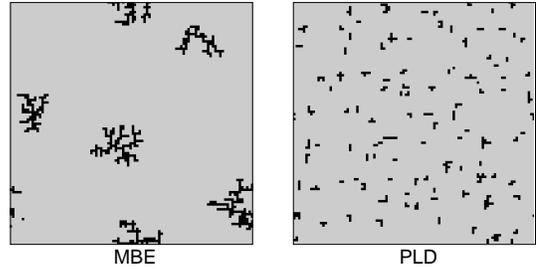, width=70mm, angle=0} 
\vspace{2mm}
\caption{Molecular beam epitaxy (left) compared to pulsed laser 
deposition (right) for $D/F=10^8$ and $I=0.01$. The figure shows
typical configurations after deposition of $0.05$ monolayers.}
\label{FigMBEPLD}
\end{center}
\end{figure}

If the intensity $I$ is very low, PLD and MBE display essentially
the same properties. However, if the intensity exceeds 
the average density of adatoms during a MBE process,
we expect a crossover to a different type of behavior. Since
this density is known to scale as $(D/F)^{1-2\gamma}$, where
$\gamma=1/(4+d_f) \simeq 1/6$, the crossover takes place at
a critical intensity 
\begin{equation}
\label{PLD_CriticalIntensity}
I_c \approx (D/F)^{-2/3} \,.
\end{equation}
The qualitative difference between PLD and MBE for $I>I_c$ is
shown in Fig.~\ref{FigMBEPLD}. As can be seen, 
there are much more nucleations at an early stage, although the 
effective flux of incoming particles is the same in both cases. 

In order to avoid the influence of the crossover at $I \approx I_c$, 
we restrict our PLD-simulations 
to a particularly simple case, namely to the limit of an infinite 
$D/F$, meaning that all adatoms nucleate or 
attach to an existing island before the next pulse arrives. 
In this limit $I_c = 0$, and the nucleation density again depends only
on two variables, $n(I,\Theta)$.

\begin{figure}
\begin{center}
\epsfxsize=85mm
\epsffile{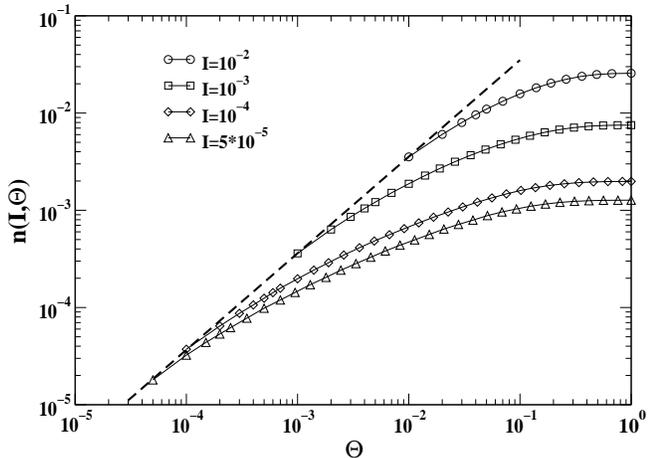}
\caption{The nucleation density versus time during the deposition 
of a monolayer. The dashed line has the slope 1.}
\label{RawResults}
\end{center}
\end{figure}

Performing Monte Carlo simulations we investigated the nucleation density
for various intensities using a system size of $400\times 400$.
The measurement always takes place right before a new pulse 
is released. Our results are shown in Fig.~\ref{RawResults}.
As can be seen, $n(I,\Theta)$ increases with increasing intensity $I$. This is
plausible since for a higher intensity more atoms arrive at the 
surface simultaneously so that most of them can meet and 
form new nucleations before attaching to already existing islands.

%
%

Obviously, the curves in Fig.~\ref{RawResults} do not display asymptotic
power laws, rather they are defined on finite intervals. 
Moreover, it is impossible to produce an ordinary data collapse
by shifting the curves horizontally and vertically in the double-logarithmic
plot. Thus, the usual scaling theory relying on power-law 
scaling fails. However, as we are going to show below, it is possible
to generate a data collapse by using the scaling form
(\ref{ApproximateScalingMBE}).

\begin{figure}
\begin{center}
\epsfxsize=85mm
\epsffile{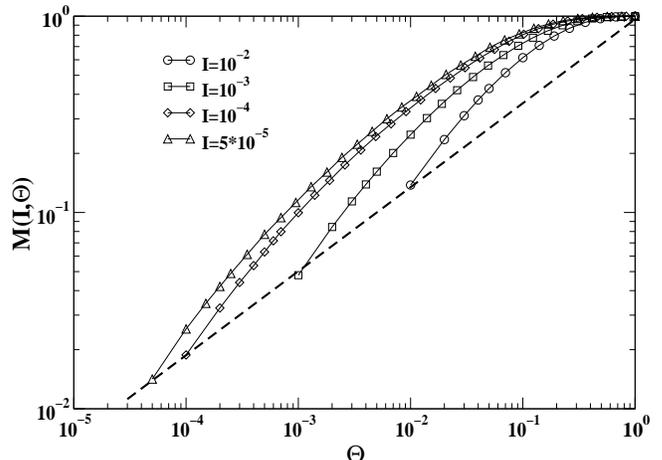}
\caption{The same data rescaled in such a way that all curves 
         terminate at the point $(1,1)$.}
\label{YScaling}
\end{center}
\end{figure}

The main idea is to stretch the curves in Fig.~\ref{RawResults}
both horizontally and vertically in such a way that their endpoints collapse.
To this end let us first consider the rightmost 
data point of each curve. As
shown in Ref.~\cite{ConferencePaper,FuturePaper} the 
saturated nucleation density $n(I,1)$ scales as 
\begin{equation}
n(I,1) \sim I^{2\nu}\,,
\label{NucleationSaturation}
\end{equation}
where the exponent $\nu$ was estimated numerically by
$\nu\simeq 0.28(1)$. The rightmost points can be collapsed by
plotting the normalized nucleation density
\begin{equation}
M(I,\Theta) = \frac{n(I,\Theta)}{n(I,1)}\,.
\label{Mcorrespondence}
\end{equation}
against $\Theta$, as shown in Fig.~(\ref{YScaling}). 

Turning to the leftmost data points in Fig.~(\ref{RawResults}), 
which correspond to the measurements after the first pulse,
we note that these points lie on a straight line with slope $1$.
This observation can be explained as follows. 
After deposition of the first pulse, 
the adatoms diffuse until they meet and nucleate. 
Most of them will nucleate with another adatom and 
form an island consisting of two atoms. Thus, after completion of the
nucleation process, the nucleation density would be $I/2$. 
In reality, however, some of the adatoms 
form bigger islands with three or more particles,
but to leading order these processes can be neglected. 
Thus, we can conclude that
\begin{equation}
n(I,I) \sim I
\label{LeftmostNucleation}
\end{equation}
which is the vertical coordinate of the leftmost data points.
Moreover, their horizontal coordinate of the points is given by 
$\Theta=I$, explaining the slope $1$ of the dashed line in Fig.~\ref{RawResults}.
In fact, the simulation data deviate from a linear law by less than $3$ $\%$.

Surprisingly the power laws~(\ref{NucleationSaturation}) 
and~(\ref{LeftmostNucleation}) are valid not only for small $I$
but also for $I \rightarrow 1$. This implies that the prefactors
of~(\ref{NucleationSaturation}) 
and~(\ref{LeftmostNucleation}) have to be identical.
This can be seen from Fig.~\ref{YScaling}, where we plotted
the normalized nucleation density $M(I,\Theta)$ versus $(\Theta)$.
In this representation the leftmost points 
of the curves obey a power law
\begin{equation}
\label{MII}
M(I,I) \sim I^{1-2\nu}
\end{equation}
extrapolating to $M(1,1)=1$, which proves the equality of the
prefactors in Eqs.~(\ref{NucleationSaturation}) 
and~(\ref{LeftmostNucleation}).

\begin{figure}
\begin{center}
\epsfxsize=85mm
\epsffile{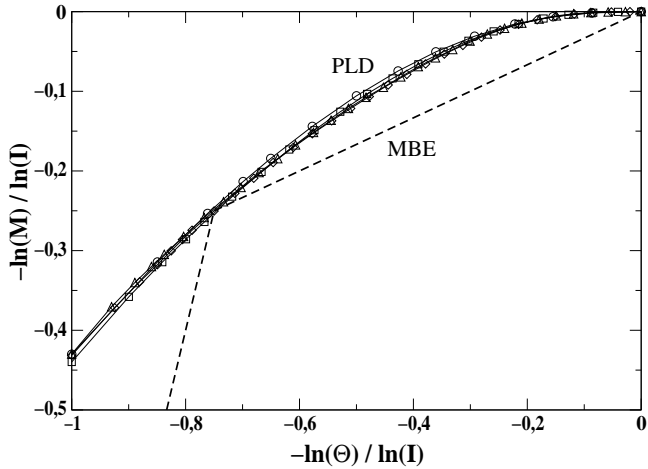}
\caption{Data collapse according to the scaling form (\ref{ScalingResult}).}
\label{FinalScaling}
\end{center}
\end{figure}

After collapsing the rightmost data points, we collapse the
leftmost data points by stretching the curves vertically
and horizontally. This can be done by dividing $\ln M$ and $\ln \Theta$
by certain factors which are proportional to $1/\ln I$. 
Clearly, this manipulation does not affect 
the upper terminal point of the curves. In fact, plotting
$\ln M/\ln I$ versus $\ln \Theta/\ln I$ (see 
Fig.~\ref{FinalScaling}), 
we obtain a convincing data collapse~\cite{Remark}.

Assuming this data collapse to hold, 
the system turns out to be invariant under
the transformation
\begin{equation}
\label{SpecificScalingScheme}
M \to M^{\lambda}\,, \qquad I \to I^\lambda\,, \qquad 
\Theta \to \Theta^{\lambda} \,.
\end{equation}
which implies the scaling form
\begin{equation}
\ln M \;\simeq\;
\left(\ln I\right) \,g\biggl(\frac{\ln \Theta}{\ln I}\biggr).
\label{ScalingResult}
\end{equation}
This scaling form has the same structure as the approximate
scaling law (\ref{ApproximateScalingMBE}) for MBE. Moreover,
we find that the scaling function is well approximated by 
a simple parabola
\begin{equation}
\label{parabola}
g(z) = a \, z^2,
\end{equation}
so that $\ln M \simeq a \ln^2\Theta/\ln I$. 
Because of the quadratic form the data collapse 
can be generated in several ways, e.g., by
plotting $\ln M$ versus $\ln \Theta/\sqrt{\ln I}$.

Finally, we present a numerical puzzle. 
As outlined before, the scaling form (\ref{ScalingResult}) for PLD 
should also hold for finite $D/F$ provided that $I \geq I_c$.
Therefore, this scaling form can be compared with the 
approximate scaling form for MBE (\ref{ApproximateScalingMBE})
at the crossover point $I \simeq I_c$, where both scaling 
concepts should `intersect'.
Because of Eqs.~(\ref{MBE_ScalingLength}) and 
(\ref{PLD_CriticalIntensity}), the system is then characterized
by the length scale $\ell_0 \sim I^{-3/8}$.
Consequently, the asympotic power laws for MBE translated into
the language of PLD read
\begin{equation}
M(I,\Theta) \simeq 
\left\{ \begin{array}{ll}
\Theta^3/I^2 & \text{ for } 0 < \Theta \ll I^{3/4} \\
\Theta^{1/3} & \text{ for } I^{3/4} \ll \Theta \ll 1
\end{array} \right.
\end{equation}
Taking the logarithm and dividing by $\ln I$ we obtain
\begin{equation}
\label{Puzzle}
\frac{\ln M(I,\Theta)}{\ln I} =
\left\{ \begin{array}{ll}
3 \frac{\ln \Theta}{\ln I}-2 & \text{ for } \frac34 \ll \frac{\ln \Theta}{\ln I}
\\[2mm]
\frac{\ln \Theta}{3 \ln I} & \text{ for } 0 \ll \frac{\ln \Theta}{\ln I} \ll \frac34
\end{array} \right.
\end{equation}
In the limit $I  \rightarrow 0$ the crossover between the two power laws
becomes sharper so that Eq.~(\ref{Puzzle}) converges to a piecewise
linear curve, which is shown in Fig.~\ref{FinalScaling}
as a dashed line. Surprisingly, the crossover point 
\begin{equation}
\label{CrossoverPoint}
\frac{\ln \Theta_c}{\ln I}  = 3/4\,, \qquad \frac{\ln M(I,\Theta_c)}{\ln I}=1/4
\end{equation}
seems to lie precisely on the collapsed curves for PLD. 
If this were true, the prefactor $a$ in Eq.~(\ref{parabola}) 
would be given by $a=4/9$. Furthermore, the
scaling relation~(\ref{MII}) implies that the prefactor 
$a$ and the exponent $\nu$ are related by $a=1-2\nu$.
Thus, we obtain the result 
\begin{equation}
\nu = (1-a)/2 = 5/18 \simeq 0.278 \, .
\end{equation}
Surprisingly, this value is in perfect agreement 
with the previous numerical estimate
$\nu \simeq 0.28(1)$ of Ref.~\cite{FuturePaper}.


To summarize, we have demonstrated that a simple model for pulsed laser
deposition displays an unusual type of scaling behavior. In its
most general setup, this kind of scaling behavior is observed in
systems which are invariant under the transformation
\begin{equation}
\label{ModifiedScalingScheme}
M \rightarrow M^{\lambda^{\alpha}}\,, 
\qquad z_i \rightarrow  z_i^{\lambda^{\beta_i}}\,,
\end{equation}
for arbitrary $\lambda$, where $\alpha$ and $\beta_1,\ldots,\beta_n$ 
are certain exponents. The corresponding scaling form reads
\begin{equation}
\label{LogarithmicScalingForm}
\ln M
\sim (\ln z_1)^\alpha \, g\biggl(\frac{\ln z_2}{(\ln z_1)^{\beta_2}} 
, \ldots ,\frac{\ln z_n}{(\ln z_1)^{\beta_n}}\biggr) \,.
\end{equation}
Performing numerical simulations, we have demonstrated that such a 
scaling form leads to a convincing data collapse for PLD in the
limit of an infinite diffusion constant. In contrast 
to ordinary scaling functions the function $g$ is defined 
on a limited interval between two points.

Various questions remain open. First of all, it would be nice
to find further examples for this kind of scaling. In this
context it would be particularly interesting to investigate 
an exactly solvable model with such a scaling behavior
and to prove the validity of the proposed scaling form. Moreover,
it is not yet clear to what extent the scaling function $g$ is
universal, i.e., independent of details of the dynamic rules of the
model. Finally, the numerical coincidence of the crossover point
(\ref{CrossoverPoint}) and a particular point of the collapsed curves
is not yet understood.


\noindent
We thank the Deutsche Forschungsgemeinschaft for
support within SFB 491.



\begin{thebibliography}{10}
\bibitem{Remark}
        In order to make the graphs comparable, we actually reflect
        the curves by plotting $-\ln M/\ln I$ versus $-\ln \Theta/\ln I$.
\end{thebibliography}
\end{document}